\documentclass[a4paper]{jpconf}
\usepackage{graphicx}
\usepackage{amssymb}
\begin{document}
\title{Prospects for CDM sub-halo detection using high angular resolution observations}

\author{T Riehm$^1$, E Zackrisson$^{2,1,3}$, O M\"oller$^4$, E M\"ortsell$^5$ and K Wiik$^2$}

\address{$^1$ Stockholm Observatory, AlbaNova University Center, SE-106 91 Stockholm, Sweden}
\address{$^2$ Tuorla Observatory, University of Turku, V\"ais\"al\"antie 20, FI-215 00 Piikki\"o, Finland}
\address{$^3$ Department of Astronomy and Space Physics, Box 515, SE-751 20 Uppsala, Sweden}
\address{$^4$ Schloss-Wolfsbrunnenweg 66, D-69118 Heidelberg, Germany}
\address{$^5$ Department of Physics, AlbaNova University Center, SE-106 91 Stockholm, Sweden}

\ead{teresa@astro.su.se}

\begin{abstract}
In the CDM scenario, dark matter halos are assembled hierarchically from smaller subunits. A long-standing problem with this picture is that the number of sub-halos predicted by CDM simulations is orders of magnitudes higher than the known number of satellite galaxies in the vicinity of the Milky Way. A plausible way out of this problem could be that the majority of these sub-halos somehow have so far evaded detection. If such ``dark galaxies'' do indeed exist, gravitational lensing may offer one of the most promising ways to detect them. 
Dark matter sub-halos in the 10$^6$ -- 10$^{10} M_{\odot}$ mass range should cause strong gravitational lensing on (sub-)milliarcsecond scales. We study the feasibility of a strong lensing detection of dark sub-halos by deriving the image separations expected for density profiles favoured by recent simulations and comparing these to the angular resolution of both existing and upcoming observational facilities. We find that there is a reasonable probability to detect sub-halo lensing effects in high resolution observations at radio wavelengths, such as produced by the upcoming VSOP-2 satellite, and thereby test the existence of dark galaxies. 
\end{abstract}

\section{Introduction}

A multitude of recent cosmological studies suggest that most of the matter in the universe consists of non-baryonic particles which are believed to constitute the cold dark matter (CDM) \cite{Komatsu}. While the CDM scenario has been very successful in explaining the formation of large-scale structures in the Universe (see e.g., \cite{Primack}, for a review), its predictions on the scales of individual galaxies have not yet been confirmed in any convincing way. One particularly interesting feature of the CDM model is the high level of halo substructure generated, seemingly inconsistent with the known number of satellite galaxies in the vicinity of the Milky Way (e.g., \cite{klypin99,moore99a,diemand07b}). This discrepancy persists even when considering the newly discovered ultra-faint dwarf galaxies in the Sloan Digital Sky Survey and correcting for its sky coverage \cite{simon07}. One possible way out of this problem is to assume that most of these low-mass halos correspond to so-called dark galaxies \cite{Verde}, i.e., objects of dwarf-galaxy mass which either do not contain baryons or in which the baryons have not formed many stars. 

Gravitational lensing may in principle offer a route to detecting even completely dark galaxies. Dark matter sub-halos in the 10$^6$--10$^{10} M_{\odot}$ range are expected to cause gravitational milli-lensing, i.e., image splitting at a characteristic separation of milliarcseconds \cite{wambsganss92,yonehara03}. To put the CDM sub-halo predictions to the test, it has been suggested that one should target quasars which are already known to be gravitationally lensed on arc second scales, as one can then be sure that there is a massive halo well-aligned with the line of sight, which substantially increases the probability for sub-halo milli-lensing \cite{yonehara03}. Indeed, the magnification associated with milli-lensing has long been suspected to be the cause of the flux ratio anomalies seen in such systems \cite{mao98,kochanek04}. Sub-halo milli-lensing has also been advocated as an explanation for strange bending angles of radio jets \cite{metcalf02} and image positions which smooth halo models seem unable to account for \cite{biggs04}.

Here, we take a critical look at the prospects for strong-lensing detections of dark sub-halos in the dwarf-galaxy mass range. In $\S$\ref{imsep}, we derive the image separations expected for sub-halo density profiles favoured by recent simulations. In $\S$\ref{lensprob}, we estimate the overall sub-halo lensing probabilities for point sources as well as extended sources. Finally, we discuss several effects that could affect our predictions and present our conclusions in $\S$\ref{concl}. 

Throughout the paper, we assume a $\Lambda$CDM cosmology with $\Omega_{\Lambda}$ = 0.762, $\Omega_{\mathrm{M}}$ = 0.238 and $h = 0.73$ ($H_0=100h$\,km\,s$^{-1}$\,Mpc$^{-1}$) in concordance with the \textit{WMAP} 3-year data release \cite{spergel}.  

\section{Image separations}
\label{imsep}

To first order, the image separation produced through strong lensing by an extended object with a density that decreases as a function of distance from the centre is given by 
\begin{equation}
\Delta\theta\approx 2 R_\mathrm{E}/D_\mathrm{ol}, 
\label{imagesepeq}
\end{equation}
where $D_\mathrm{ol}$ represents the angular-size distance between observer and lens, and  $R_\mathrm{E}$ represents the linear Einstein radius. The latter is defined as the radius inside which the mean surface mass density $\bar{\Sigma}$ of the lens equals the critical surface mass density 
\begin{equation}
\bar{\Sigma}(<R_\mathrm{E})=\Sigma_\mathrm{c}=\frac{c^2D_\mathrm{os}}{4\pi G D_\mathrm{ol}D_\mathrm{ls}},
\end{equation}
where $D_\mathrm{os}$ and $D_\mathrm{ls}$ are the angular-size distances between observer and source, and lens and source, respectively.

The resulting image separations will thus heavily depend on the sub-halo surface mass density profiles $\Sigma(r)$. Previously proposed strategies to detect image splitting \cite{yonehara03,inoue05a} assume that the image separations of sub-halo lenses are similar to those produced by a singular isothermal sphere (SIS). Unfortunately, this assumption is difficult to justify since theoretical arguments, simulations, and observations do not favour this form of density profile for dark matter halos in the relevant mass range. We have studied the feasibility of strong-lensing detection of dark sub-halos by deriving the image separations expected for (more realistic) density profiles favoured by recent simulations. These are NFW \cite{nfw}, M99 \cite{moore99b}, N04 \cite{Navarro04}, H03 \cite{Hayashi} and K04 \cite{Kazantzidis} where we also have taken the impact of stripping and truncation into account. Details on these density profiles and a comparison of the resulting mean surface mass density profiles $\bar{\Sigma}(<r)$, can be found in \cite{ez}.

\begin{figure}[t]
\includegraphics[clip,width=0.48\textwidth]{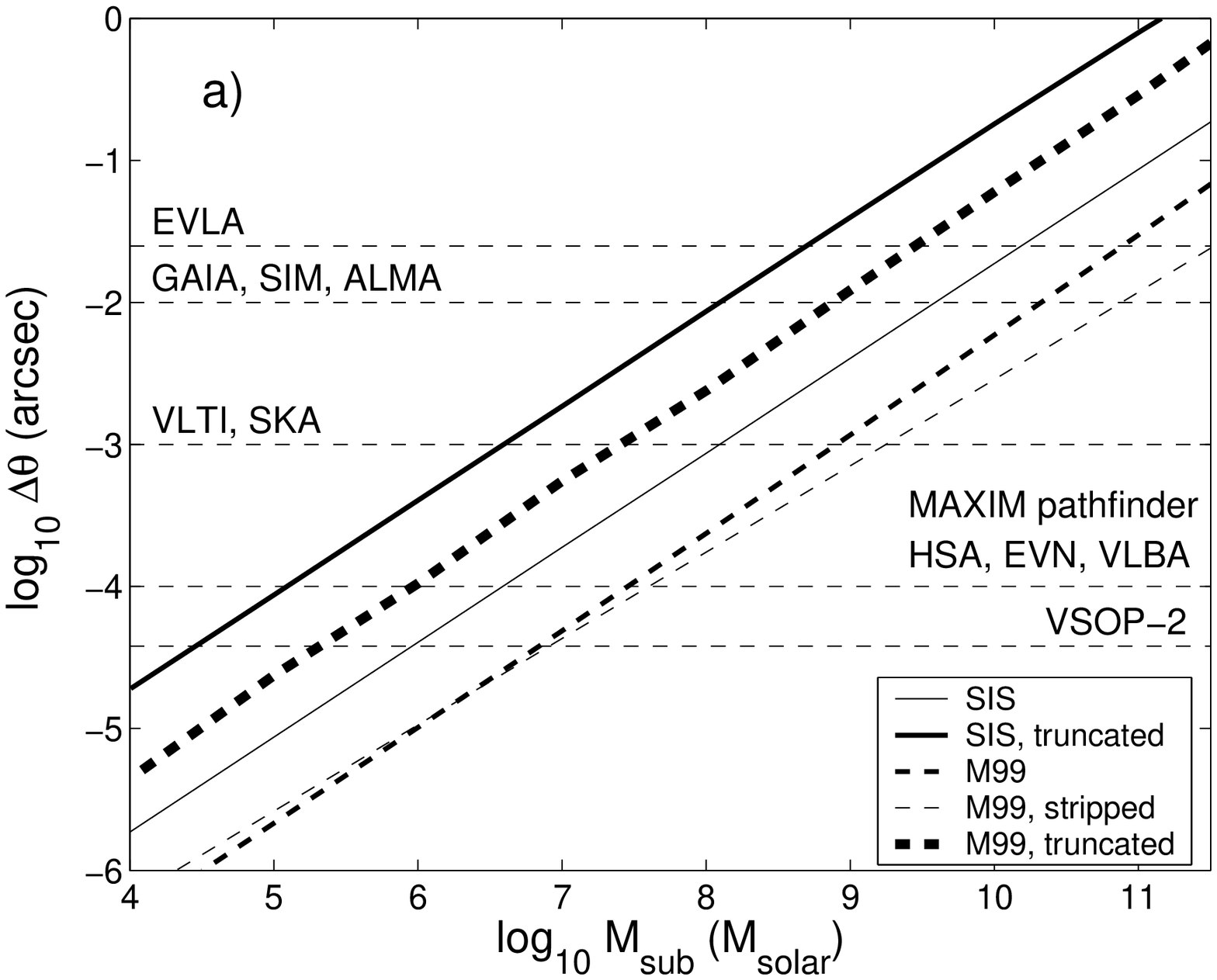}\hfill%
\includegraphics[clip,width=0.48\textwidth]{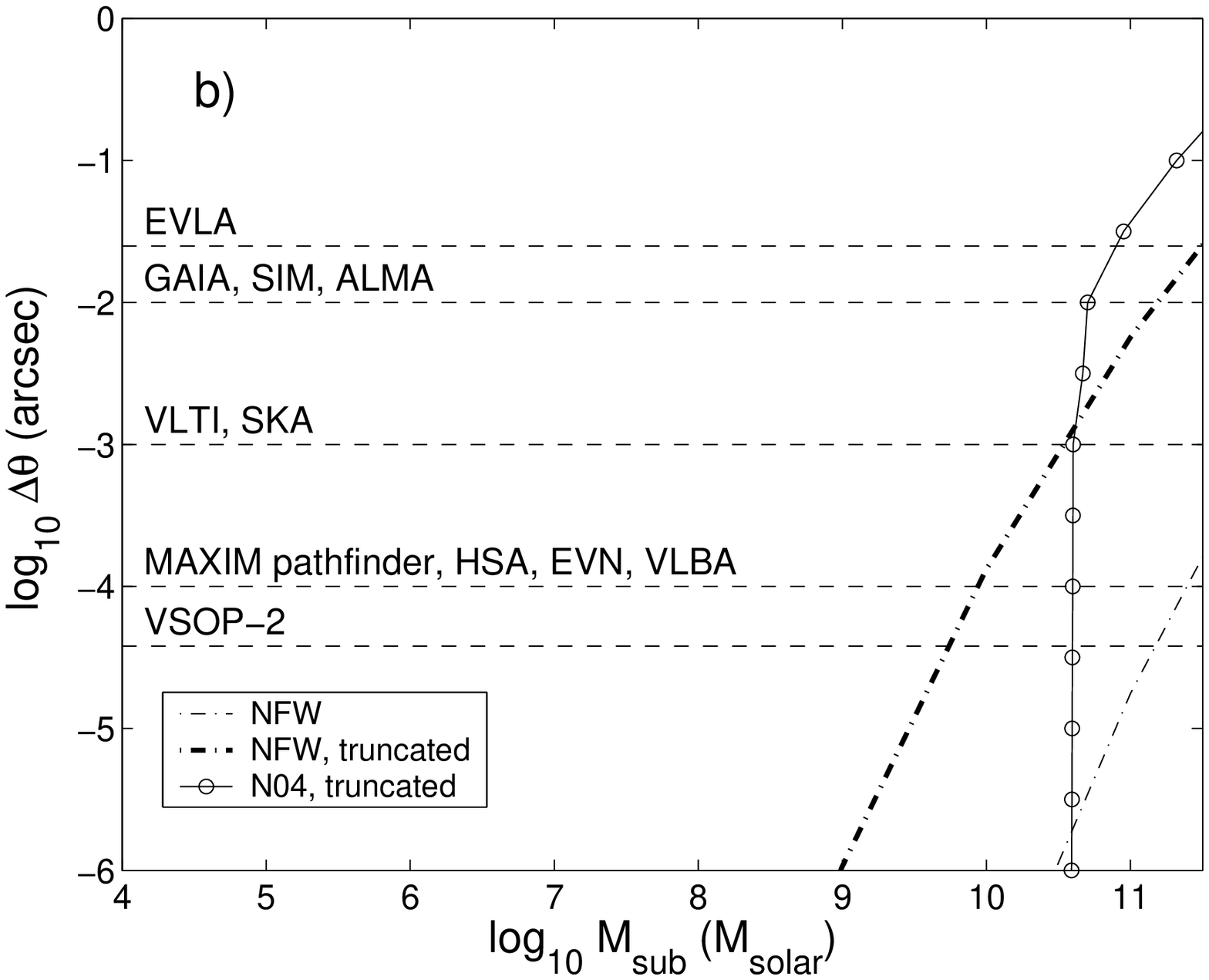}
\caption{\label{fig1} Sub-halo mass versus image separation $\Delta\theta$ for those density profiles that give rise to image separations on scales larger than micro-arc seconds. The angular resolution of a number of existing and planned observational facilities have been indicated by horizontal dashed lines, marked with labels (see main text for details). a) The different diagonal lines represent SIS (thin solid), truncated SIS (thick solid), M99 (medium dashed), stripped M99 (thin dashed) and truncated M99 (thick dashed) sub-halo models. b) The different diagonal lines represent NFW (thin dash-dotted), truncated NFW (thick dash-dotted) and truncated N04 (solid with circles) sub-halo models.}
\end{figure}

In Fig.~\ref{fig1}, we plot the image separations predicted for $10^4$--$10^{11}\ M_\odot$ sub-halos at a redshift $z_{\mathrm{l}} = 0.5$ and a source at $z_{\mathrm{s}} = 2.0$. We compare this to the angular resolution of a number of planned or existing observational facilities, operating at a wide range of wavelengths. These include the proposed \textit{MAXIM} pathfinder\footnote{maxim.gsfc.nasa.gov/pathfinder.html} in X-rays; VLTI with the proposed VSI instrument \cite{Malbet}, the planned \textit{GAIA}\footnote{www.rssd.esa.int/index.php?project=GAIA} and \textit{SIM PlanetQuest}\footnote{planetquest.jpl.nasa.gov/SIM/} satellites in the optical/near-infrared; the currently available EVN\footnote{www.evlbi.org}, HSA\footnote{www.nrao.edu/HSA/}, VLBA\footnote{www.vlba.nrao.edu} arrays plus the planned ALMA\footnote{www.alma.info}, EVLA\footnote{www.aoc.nrao.edu/evla}, SKA\footnote{www.skatelescope.org} arrays, and also space-VLBI with the planned VSOP-2\footnote{www.vsop.isas.ac.jp/vsop2} programme at radio wavelengths. Please note that here we consider only the best resolution limits attainable with these telescopes, whereas the resolution at the wavelengths that maximise the number of observable high-redshift sources may be considerably worse. 

From Fig.~\ref{fig1} it becomes obvious that there are large differences between the image separation predictions of the various halo models. As the discrepancy between the number densities of luminous galaxies and dark matter halos does not start to become severe until the halo mass drops below $10^{10} \ M_\odot$ (e.g., \cite{Verde,vandenBoscha}), sub-halos at masses below this limit need to produce measurable image separations ($\theta \gtrsim 4\times 10^{-5}$$^{\prime\prime}$ for VSOP-2, which has the best theoretical resolution among the telescopes included in Fig.~\ref{fig1}) in order for dark galaxies to be detectable through image-splitting effects. Out of the halo models tested, only two actually meet this criterion without adhering to sharp truncations: the SIS and the M99 halos. The H03 and K04 profiles both give image separations smaller than $10^{-6}$$^{\prime\prime}$ for all the halo masses considered and are therefore completely outside the plotted region. Even in the optimistic case of an M99 halo (in either its original or stripped form, whereas the sharp truncation is not considered realistic, see \cite{ez} for details), the image separations are a factor of $\approx 3$--7 smaller than those predicted for a SIS (and $\approx30$--60 times smaller than those of a truncated SIS), rendering only the few most massive sub-halos ($\sim 10^{10} \ M_\odot$ or slightly higher) detectable at $\sim0.01^{\prime\prime}$ resolution (\textit{GAIA, SIM}, and ALMA). At milliarcsecond resolution (VLTI and SKA), dark galaxies with masses $\gtrsim10^9 \ M_\odot$ may become detectable. To probe further down the sub-halo mass function, sub-milliarcsecond-resolution facilities (\textit{MAXIM pathfinder}, HSA, EVN, VLBA or VSOP-2) will be required. 

These estimates are based on the assumption that the sub-halos can be treated as isolated objects. The effects of external convergence $\kappa$ and shear $\gamma$ due to the galaxy (and halo) hosting these sub-halos have thereby been neglected. The question remains whether the boost factor $f_\mathrm{boost}=\Delta\theta_{\kappa,\gamma}/\Delta\theta_{\kappa=0,\gamma=0}$ from the host potential can become considerably large for some of the more realistic density profiles considered here. To investigate this, we use ray-tracing simulations to numerically assess the distribution of $f_\mathrm{boost}$. We concentrate on two sub-halo models that span the range from effectively undetectable (H03) to favourable for detection (M99). In both cases, the tail of the $f_\mathrm{boost}$ distribution extends up to very high values, which means that it is in principle possible to find macro-images for which sub-halos of masses much lower than indicated by Fig.~\ref{fig1} can be detected through image-splitting effects. However, such macro-images are exceedingly rare. For sub-halos with  $M_\mathrm{sub}=10^{10} \ M_\odot$, the probability of having $f_\mathrm{boost}\geq 100$ is $\sim 10^{-3}$ in the case of M99 and $\approx 0.03$ in the case of H03. The expectation value for the boost factor is $\langle f_\mathrm{boost}\rangle\approx 2.3$ for the M99 sub-halo and $\langle f_\mathrm{boost}\rangle\approx 14$ for the H03 sub-halo. This is enough to shift the $\Delta\theta$ prediction for the M99 sub-halos fairly close to that of an un-truncated SIS in Fig.~\ref{fig1}, but insufficient to move the H03 prediction ($\Delta\theta_{\kappa=0,\gamma=0} \sim 10^{-15}$$^{\prime\prime}$ at this mass) into the detectable range.

\section{Lensing probabilities}
\label{lensprob}

We assume a sub-halo population following the model proposed in \cite{gao04}. The sub-halo abundance per unit host halo mass (ignoring the high mass cut-off, $M_{\mathrm{sub}} > 0.01 M_{\mathrm{host}}$) can be evaluated by
\begin{equation}
\frac{dM_{\mathrm{sub}}}{dm} = 10^{-3.2} \left(\frac{M_{\mathrm{sub}}}{h^{-1} M_{\odot}}\right)^{-1.9} h M_{\odot}^{-1} .
\end{equation}

In \cite{diemand07a}, this finding was confirmed and extended to lower sub-halo masses using a high resolution simulation of CDM substructure in a Milky Way-sized halo. Within a host halo there is then a fraction $f_{\mathrm{sub}}(M_{\mathrm{host}},M_{\mathrm{sub\_min}},M_{\mathrm{sub\_max}})$ of its mass $M_{\mathrm{host}}$ in the form of sub-halos with minimum and maximum masses $M_{\mathrm{sub\_min}}$ and $M_{\mathrm{sub\_max}}$, respectively. Here we assume sub-halo masses $M_{\mathrm{sub}}$ in the range $4 \times 10^{6} \leq M_{\mathrm{sub}}/M_\odot \leq 10^{10}$, corresponding to the interval probed in \cite{diemand07a}. The total mass fraction of a host halo with $M$ = $1.8 \times 10^{12} M_{\odot}$ in sub-halos is then about 5$\%$ for this sub-halo mass range.  

Within this simulation, it has also been shown that the sub-halo number density profile can be fitted by the following form \cite{madau08}:
\begin{equation}
\frac{n(<x)}{N} = \frac{12x^3}{1+11x^2},
\end{equation}
where $x$ is the distance to the host center in units of $R_{200}$, the radius at which the mean enclosed density equals 200 times the mean mass density of the Universe, $n(<x)$ is the number of sub-halos within $x$ and $N$ is the total number of sub-halos inside $R_{200}$.

\subsection{Point sources}

Under the assumption that the lenses do not overlap along the line of sight, the optical depth $\tau$ represents the fraction of a given patch of the sky that is covered by regions in which a point source will be lensed. 
In the limit of small $\tau$, the optical depth can directly be used as an estimate of the lensing probability. 

\begin{equation}
\tau(\xi,\kappa, \gamma) = \frac{1}{S}\int_{M_{\mathrm{sub\_min}}} ^{M_{\mathrm{sub\_max}}} \sigma_{\mathrm{lens}}(M_{\mathrm{sub}},\kappa,\gamma) \frac{dn(\xi)}{dM_{\mathrm{sub}}} dM_{\mathrm{sub}}
\end{equation}
with the area of a patch on the sky $S$, the projected radius from the center of the host halo $\xi$ and the external convergence $\kappa$ and shear $\gamma$ induced by the host halo. Here $\sigma_{\mathrm{lens}}$ denotes the cross section for a single sub-halo lens  within the potential of the host halo. We use the analytic expression derived in \cite{keeton03} to compute the lensing cross section $\sigma_{\mathrm{lens}}$ for SIS sub-halos. 

\begin{figure}[t]
\includegraphics[clip,width=0.65\textwidth]{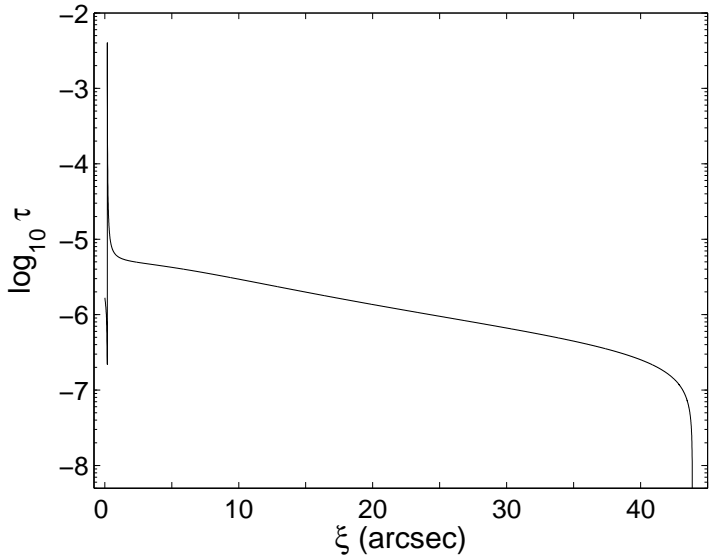}\hfill%
\begin{minipage}[b]{0.33\textwidth}\caption{\label{fig2} Optical depth for a point source at $z_{\mathrm{s}}$ = 1 as a function of projected radius. Here we assume a host halo at $z_{\mathrm{l}}$ = 0.5 of mass $M_{\mathrm{host}} = 1.8 \times 10^{12}$ $M_{\odot}$ and sub-halos in the mass range $4 \times 10^{6}$ -- 10$^{10}$ $M_{\odot}$. The peak has been cut with respect to a maximum magnification factor of 50 at the Einstein radius of the host halo.}
\end{minipage} 
\end{figure}

As can be seen in Fig. \ref{fig2}, the optical depth for a typical scenario does not exceed a value of $4 \times 10^{-3}$ at any radius. This is contrary to prior claims where the optical depth for sub-halo lensing had been estimated to be several orders of magnitude higher \cite{yonehara03}. This disagreement can be traced to an error in the sub-halo mass function adopted by the previous study.

\subsection{Extended sources}

According to \cite{inoue05b}, it should be possible to detect lensing effects from dark matter substructures when observing extended sources resolved at scales smaller than the Einstein radii of the sub-halos. Furthermore, one should even be able to put constraints on the internal density profiles of the lensing sub-halos. This technique may already become observationally feasible with ALMA \cite{inoue05a} or future space-VLBI missions like VSOP-2 \cite{inoue05c} and thus constitute a major step forward in the study of dark halo substructures. Here, we estimate the probability of sub-halo lensing for e.g. a quasar in the radio regime by computing the average number of sub-halos that lie within the region of the host halo covering the source. 

\begin{figure}[t]
\includegraphics[clip,width=0.60\textwidth]{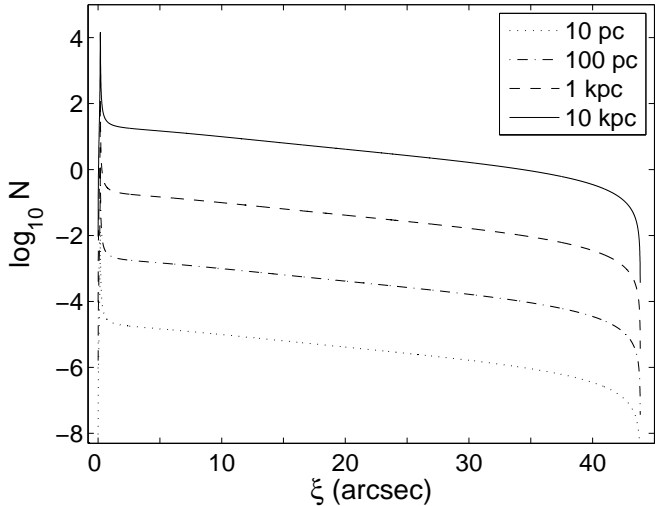}\hfill%
\begin{minipage}[b]{0.38\textwidth}\caption{\label{fig3} Average number of substructures covering an extended source at $z_{\mathrm{s}}$ = 1 as a function of projected radius for different source sizes. We assume a host at $z_{\mathrm{l}}$ = 0.5 with $M_{\mathrm{host}}$ = $1.8 \times 10^{12} M_{\odot}$ and sub-halos in the range $4 \times 10^{6}$ -- $10^{10} M_{\odot}$. We plot our results for a source with radius $r_{\mathrm{s}}$ = 10 pc (dotted line), 100 pc (dash-dotted line), 1\,kpc (dashed line) and 10\,kpc (solid line), respectively. The peak has been cut with respect to a maximum magnification factor of 30 at the Einstein radius of the host.}
\end{minipage} 
\end{figure}

In Fig. \ref{fig3}, the expected number of sub-halos covering an extended source at $z_{\mathrm{s}}$ = 1 as a function of projected radius from the host halo lens center is shown for several source radii $r_{\mathrm{s}}$, ranging from 10 pc to 10\,kpc. This can be compared to the virial radius $R_{200} \approx$260\, kpc for the host galaxy at $z_{\mathrm{l}}$ = 0.5. It becomes clear that for sufficient source size ($\gtrsim$1\,kpc) there is a good probability for the source image to be affected by sub-halo lensing, not only close to the Einstein radius of the host halo but even at a rather large projected distance from the host halo lens center. For a source with $r_{\mathrm{s}}=1$\,kpc, one would expect at least one intervening sub-halo per 10 observed systems with a maximum projected distance of 10$^{\prime\prime}$ between the foreground galaxy and the source. For $r_{\mathrm{s}} = 10$\,kpc, this number increases to approximately 10 sub-halos projected on the source out to a distance of 10$^{\prime\prime}$ from the host galaxy.

\begin{figure}[t]
\includegraphics[clip,width=0.60\textwidth]{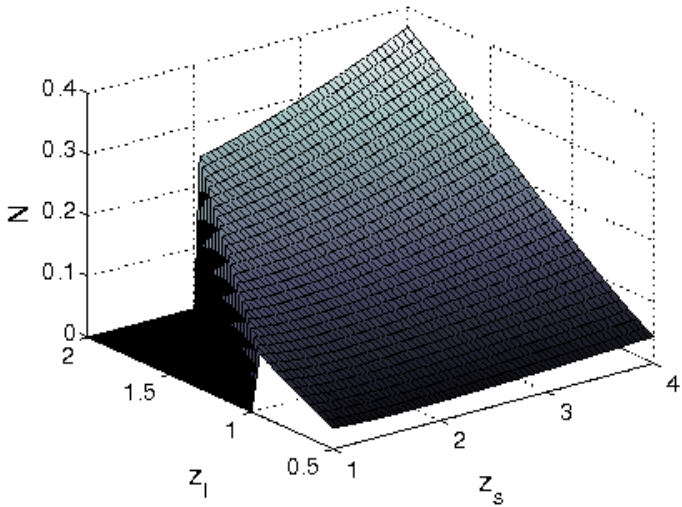}\hfill%
\begin{minipage}[b]{0.38\textwidth}\caption{\label{fig4} Average number of substructures covering a source with radius $r_{\mathrm{s}} = 1$\,kpc at a projected radius of $R_{200}/2$ (ranging from 22$^{\prime\prime}$--7$^{\prime\prime}$ for $z_{\mathrm{l}}$ = 0.5 -- 2) as a function of lens redshift, $z_{\mathrm{l}}$, and source redshift, $z_{\mathrm{s}}$. We assume a host halo of mass $M = 1.8 \times 10^{12} M_{\odot}$ and sub-halos in the mass range $4 \times 10^{6}$ -- $10^{10} M_{\odot}$.}
\end{minipage} 
\end{figure}

We also explore the dependence of these results on the lens and source redshifts, $z_{\mathrm{l}}$ and $z_{\mathrm{s}}$ respectively. The probability for a quasar to be aligned close to the Einstein radius of a foreground galaxy, where its magnification will boost the expected number of intervening sub-halo lenses, is low even when considering magnification bias. However, it is expected that there is a large number of galaxy-quasar pairs with a projected separation smaller than the virial radius of the host halo $R_{200}$ (typically up to a few tens of arc seconds). In Fig. \ref{fig4}, the average number of intervening sub-halos at $R_{200}/2$ for a  source size of $r_{\mathrm{s}} = 1$\,kpc is shown as a function of $z_{\mathrm{l}}$ and $z_{\mathrm{s}}$. The probability for lensing by sub-halos increases by up to one order of magnitude for high-$z$ objects. 

Since high-mass host halos typically possess a larger fraction of their mass clustered in substructure, this will also boost the expected number of sub-halos projected on the source. Furthermore, high-mass host halos possess larger virial radii $R_{200}$, allowing for larger projected distances between the host galaxy and the background source (see \cite{Riehm} for details).

\section{Discussion and Conclusions}
\label{concl}

Our results indicate that the detection of sub-halos through gravitational image-splitting is likely to be considerably more challenging than suggested in previous studies, due to the smaller image separations predicted for sub-halo density profiles more realistic than the SIS models often adopted. In fact, no currently planned telescope will be able to resolve the image separations produced by sub-halos with density profiles of the type suggested by the most realistic simulations recently available (H03 \& K04). If the sub-halos would have steeper central density slopes (e.g., of M99 type), these would give rise to image separations that could be resolved even with existing telescopes. A recent simulation with the highest resolution currently available \cite{diemand08}, obtained central density profiles of the form $\rho(r) \propto r^{-\alpha}$ with central density slopes $\alpha \approx 1.24$ (i.e., between the unfavourable NFW with $\alpha = 1$ and the optimistic M99 profile with $\alpha = 1.5$). If strong lensing effects produced by sub-halos which such profiles could be observed with any present or future telescope, is something we will be looking into in the near future. 

Despite the somewhat bleak detection prospects presented here, there are at least two effects that can potentially improve the detectability of image splitting by sub-halos: baryon cooling and the presence of intermediate mass black holes. 

Baryon cooling would cause the sub-halo (or its progenitor halo) to contract (e.g., \cite{Gnedin,Maccio,Gustafsson,Kampakoglou}), thereby increasing the central density and boosting the image separation. It is, however, not clear how strong this effect is likely to be, since this depends on the details of the mechanism that prevents the dark sub-halo from forming stars. Baryon cooling is usually assumed to be associated with star formation, and most attempts to explain the lack of bright sub-halos therefore propose that the baryons have either been lost early in the history of the Universe, or are prevented from cooling by the ultraviolet background provided by re-ionization \cite{Barkana,Read}. Neither mechanism is likely to result in any significant halo contraction. Nonetheless, claims of dark baryons in the form of cold gas in galactic disks have been made \cite{Bournaud}, implying that there may be some route for gas to cool without forming stars or being detected by the usual H$_2$ tracers \cite{Pfennigera,Pfennigerb}. One may therefore speculate that there could be alternative scenarios, in which sub-halos are kept dark even though baryon cooling has taken place.

Intermediate mass black holes (IMBHs, see e.g., \cite{vanderMarel,Zhao,Noyola}) with masses of $\sim 10^2$--$10^4\ M_\odot$ would also boost the image separations, if present in the centres of dark sub-halos. This could for instance be the case if the empirical relations between the mass of a super-massive black hole and its dark matter halo \cite{Ferrarese} would extend into the dwarf-galaxy mass range. Even in the case of $\kappa=\gamma=0$, an IMBH of mass $10^4 \ M_\odot$, would give an image separation of $\approx 4\times 10^{-4}$$^{\prime\prime}$ for $z_\mathrm{l}=0.5$ and $z_\mathrm{s}=2.0$. This is sufficient to allow VSOP-2 to resolve the image splitting, regardless of the density profile of the sub-halo hosting the IMBH.  Of course, even if the Ferrarese relations would hold for luminous dwarf galaxies, they do not necessarily do so for dark ones, since this depends on the formation details of IMBHs. Moreover, a population of halo IMBHs {\it not} associated with sub-halos could form an undesired background of milli-lensing events that would obfuscate attempts to study sub-halos through image-splitting effects.

We have shown that the optical depth $\tau$ for sub-halo lensing of point sources is lower than previously predicted. Even for the most favorable conditions it barely exceeds 0.01. We conclude that it is currently not feasible to use this technique to search for strong lensing signatures in point sources as e.g. quasars in the optical.

If one instead targets extended sources, such as quasars in the radio wavelength regime, there is a high probability for sub-halo lensing of sources of sufficient size. For source sizes $r_{\mathrm{s}} \gtrsim 1$\,kpc, this is valid even at rather large projected distance of the source to the host halo center. This allows for a different search strategy than those previously proposed. Instead of only targeting multiply-imaged quasar systems, even quasar-galaxy pairs with a separation of several tens of arc seconds should show effects of strong lensing by substructures in the lens galaxy halo.

\begin{figure}[t]
\includegraphics[clip,width=0.60\textwidth]{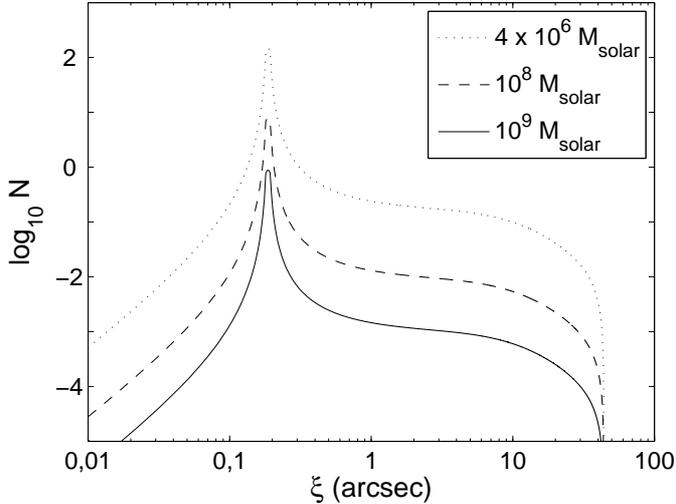}\hfill%
\begin{minipage}[b]{0.38\textwidth}\caption{\label{fig5} Average number of substructures covering a source with $r_{\mathrm{s}} = 1$\,kpc at $z_{\mathrm{s}} = 1$ as a function of projected radius for different minimum sub-halo masses $M_{\mathrm{sub\_min}}$. For a host halo at $z_{\mathrm{l}} = 0.5$ with $M_{\mathrm{host}} = 1.8 \times 10^{12} M_{\odot}$, we vary $M_{\mathrm{sub\_min}}$ with $4 \times 10^6 M_{\odot}$ (dotted line), $10^8 M_{\odot}$ (dashed line) and $10^9 M_{\odot}$ (solid line), respectively. The peak has been cut with respect to a maximum magnification factor of 30 at the Einstein radius of the host halo.}
\end{minipage} 
\end{figure}

However, these effects will strongly depend on the sub-halo mass and density profile. Even for the most favorable density profiles, there will be minimum masses for which the image separation drops below the resolution of any present or planned observational facility. E.g., for the currently available EVN, the minimum sub-halo mass which could be resolved under optimal conditions is approximately $4 \times 10^6 M_{\odot}$ in case of an SIS profile but increases to $3 \times 10^7 M_{\odot}$ for an M99 profile. Figure \ref{fig5} shows how the expected number of intervening substructures $N$ depends on the minimum sub-halo mass $M_{\mathrm{sub\_min}}$. For a host halo of mass $M = 1.8 \times 10^{12} M_{\odot}$ at $z_{\mathrm{l}}$ = 0.5 and a source of size $r_{\mathrm{s}} = 1$\,kpc at $z_{\mathrm{s}}$ = 1, we compute $N$ as a function of projected radius from the host halo center for minimum sub-halo mass $M_{\mathrm{sub\_min}} = 4 \times 10^6 M_{\odot}$, $10^8 M_{\odot}$ and $10^9 M_{\odot}$, respectively. Since the sub-halo mass function predicts about equal mass within each logarithmic mass interval, most sub-halos will be of low mass ($n(M_{\mathrm{sub}}) \propto M_{\mathrm{sub}}^{-0.9}$). Thus, the expected number of intervening substructures is very sensitive to the minimum sub-halo mass that can be resolved. 

Therefore, angular resolution will be crucial when attempting to detect CDM substructure via its lensing effects on background quasars and sub-milliarcsecond-resolution facilities will be required. We assess that the prospects for such a detection with the upcoming VSOP-2 satellite are good, with the condition that the ground array
should be extensive enough. This is not a problem especially at 8.4 GHz where a number of large antennas are available from the geodetic VLBI network. However, one must be careful not to confuse internal structures found in quasars observed at radio wavelengths with sub-halo lensing signals. Quasars already macro-lensed on arc second scales can be used to test that it is possible to distinguish between the two, since internal structures should be mapped in all of the macro-images while sub-halo lensing will only affect one of the images. Taking the above into account, one should be able to use this technique to put constraints on dark matter sub-halos predicted by simulations in the near future.  

\ack
TR acknowledges support from the HEAC Centre funded by the Swedish Research Council. EZ acknowledges research grants from the Swedish
Research Council, the Royal Swedish Academy of Sciences and the Academy of Finland.


\section*{References}
\begin{thebibliography}{9}
\bibitem{Komatsu} Komatsu E \textit{et al} 2008 Five-Year Wilkinson Microwave Anisotropy Probe (WMAP) Observations: Cosmological Interpretation ({\it Preprint} 0803.0547 {astro-ph}) 
\bibitem{Primack} Primack J R 2003, {\it Nucl. Phys.} B Proceedings Supplements {\bf 124} 3
\bibitem{klypin99} Klypin A, Kravtsov A V, Valenzuela O and Prada F 1999 {\it Astrophys. J.} {\bf 522} 82
\bibitem{moore99a} Moore B, Ghigna S, Governato F, Lake G, Quinn T, Stadel J and Tozzi P 1999 {\it Astrophys. J.} {\bf 524} L19 
\bibitem{diemand07b} Diemand J, Kuhlen M and Madau P 2007 {\it Astrophys. J.} {\bf 667} 859 
\bibitem{simon07} Simon J D and Geha M 2007 {\it Astrophys. J.} {\bf 670} 313
\bibitem{Verde} Verde L, Oh S P and Jimenez, R 2002 {\it Mon. Not. R. Astron. Soc.} {\bf 336} 541
\bibitem{wambsganss92} Wambsganss J and Paczynski B 1992 {\it Astrophys. J. Lett.} {\bf 397} L1
\bibitem{yonehara03} Yonehara A, Umemura M and Susa H 2003 {\it PASJ} {\bf 55} 1059
\bibitem{mao98} Mao S and Schneider P 1998 {\it Mon. Not. R. Astron. Soc.} {\bf 295} 587
\bibitem{kochanek04} Kochanek C S and Dalal N 2004 {\it Astrophys. J.} {\bf 610} 69
\bibitem{metcalf02} Metcalf R B 2002 {\it Astrophys. J.} {\bf 580} 696
\bibitem{biggs04} Biggs A D, Browne I W A, Jackson N J, York T, Norbury M A, McKean J P 
    and Phillips P M 2004 {\it Mon. Not. R. Astron. Soc.} {\bf 350} 949
\bibitem{spergel} Spergel D N \textit{et al} 2007 {\it Astrophys. J. Suppl. Series} {\bf 170} 377     
\bibitem{inoue05a} Inoue K T and Chiba M 2005 {\it Astrophys. J.} {\bf 633} 23
\bibitem{nfw} Navarro J F, Frenk C S and White S D M 1997 {\it Astrophys. J.} {\bf 490} 493 (NFW)
\bibitem{moore99b} Moore B, Quinn T, Governato F, Stadel J and Lake G 1999 {\it Mon. Not. R. Astron. Soc.} {\bf 310} 1147 (M99)
\bibitem{Navarro04} Navarro J F, Hayashi E, Power C, Jenkins A R, Frenk C S, White S D M, Springel V, Stadel J and Quinn T R 2004 {\it Mon. Not. R. Astron. Soc.} {\bf 349} 1039 (N04)
\bibitem{Hayashi} Hayashi E, Navarro J F, Taylor J E, Stadel J and Quinn T 2003 {\it Astrophys. J.} {\bf 584} 541 (H03)
\bibitem{Kazantzidis} Kazantzidis S, Mayer L, Mastropietro C, Diemand J, Stadel J and Moore B 2004 {\it Astrophys. J.} {\bf 608} 663 (K04)

\bibitem{ez} Zackrisson E, Riehm T, M\"oller K and Wiik K 2008 {\it Astrophys. J.} {\bf 684} 804
\bibitem{Malbet} Malbet F \textit{et al} 2008 {\it Preprint} 0801.2694 {astro-ph} 
\bibitem{vandenBoscha} van den Bosch F C, Yang X and Mo H J 2003 {\it Mon. Not. R. Astron. Soc.} {\bf 340} 771

\bibitem{gao04} Gao L, White S D M, Jenkins A, Stoehr F and Springel V 2004 {\it Mon. Not. R. Astron. Soc.} {\bf 355} 819 
\bibitem{diemand07a} Diemand J, Kuhlen M and Madau P 2007 {\it Astrophys. J.} {\bf 657} 262 
\bibitem{madau08} Madau P, Diemand J and Kuhlen M 2008 {\it Astrophys. J.} {\bf 679} 1260 
\bibitem{keeton03} Keeton C R 2003 {\it Astrophys. J.} {\bf 584} 664
\bibitem{inoue05b} Inoue K T and Chiba M 2005 {\it Astrophys. J.} {\bf 634} 77
\bibitem{inoue05c} Inoue K T and Chiba M 2005 ({\it Preprint} 0512648 {astro-ph})
\bibitem{Riehm} Riehm T, Zackrisson E, M\"ortsell E and Wiik K 2008 {\it Astrophys. J.} submitted
\bibitem{diemand08} Diemand J, Kuhlen M, Madau P, Zemp M, Moore B, Potter D and Stadel J 2008 {\it Nature} {\bf 454} 735
\bibitem{Gnedin} Gnedin O Y, Kravtsov A V, Klypin A A and Nagai D 2004 {\it Astrophys. J.} {\bf 616} 16
\bibitem{Maccio} Macci\`o A V, Moore B, Stadel J and Diemand J 2006 {\it Mon. Not. R. Astron. Soc.} {\bf 366} 1529 
\bibitem{Gustafsson} Gustafsson M, Fairbairn M and Sommer-Larsen J 2006 {\it Phys. Rev.} D 74l3522 
\bibitem{Kampakoglou} Kampakoglou M 2006 {\it Mon. Not. R. Astron. Soc.} {\bf 369} 1988
\bibitem{Barkana} Barkana R and Loeb A 1999 {\it Astrophys. J.} {\bf 523} 54
\bibitem{Read} Read J I, Pontzen A P and Viel M 2006 {\it Mon. Not. R. Astron. Soc.} {\bf 371} 885 
\bibitem{Bournaud} Bournaud F, Duc P-A, Brinks E, Boquien M, Amram P, Lisenfeld U, Koribalski B S, Walter F and Charmandaris V 2007 {\it Science} {\bf 316} 1166
\bibitem{Pfennigera} Pfenniger D, Combes F and Martinet L 1994 {\it Astron. \& Astrophys.} {\bf 285} 79
\bibitem{Pfennigerb} Pfenniger D and Combes F. 1994 {\it Astron. \& Astrophys.} {\bf 285} 94
\bibitem{vanderMarel} van der Marel R P 2004 {\it Coevolution of Black Holes and Galaxies} ed L C Ho (Cambridge: Cambridge University Press) p 37 ({\it Preprint} astro-ph/0302101)
\bibitem{Zhao} Zhao H and Silk J 2005 {\it Phys. Rev. Lett.} {\bf 95} 011301
\bibitem{Noyola} Noyola E, Gebhardt K and Bergmann M 2008 {\it Astrophys. J.} {\bf 676} 1008
\bibitem{Ferrarese} Ferrarese L 2002 {\it Astrophys. J.} {\bf 578}, 90 
\end{thebibliography}
\end{document}